# The Effect of Mechanical Strain on Lithium Staging in Graphene

**Authors:** Joshua V. Pondick[1,2], Sajad Yazdani[1,2], Milad Yarali[1,2], Serrae N. Reed[1,2], David J. Hynek[1,2], Judy J. Cha[1,2]*

**Affiliations:**

[1]Department of Mechanical Engineering and Materials Science, Yale University, New Haven, Connecticut 06511, United States.

[2]Energy Sciences Institute, Yale West Campus, West Haven, Connecticut 06516, United States.

*Correspondence to: judy.cha@yale.edu

Lithium intercalation into graphite is the foundation for the lithium-ion battery[1,2], and the thermodynamics of the lithiation of graphitic electrodes have been heavily investigated[3]. Intercalated lithium in bulk graphite undergoes structural ordering known as staging to minimize electrostatic repulsions within the crystal lattice[4]. While this process is well-understood for bulk graphite, confinement effects become important at the nanoscale, which can significantly impact the electrochemistry of nanostructured electrodes[5,6]. Therefore, graphene offers a model platform to study intercalation dynamics at the nanoscale by combining on-chip device fabrication and electrochemical intercalation with *in situ* characterization[6–8]. We show that microscale mechanical strain significantly affects the formation of ordered lithium phases in graphene. *In situ* Raman spectroscopy of graphene microflakes mechanically constrained at the edge during lithium intercalation reveals a thickness-dependent increase of up to 1.26 V in the electrochemical potential that induces lithium staging. While the induced mechanical strain energy increases with graphene thickness to the fourth power, its magnitude is small compared to the observed increase in electrochemical energy. We hypothesize that the mechanical strain energy increases a nucleation barrier for lithium staging, greatly delaying the formation of ordered lithium phases. Our results indicate that electrode assembly can critically impact lithium staging dynamics important for cycling rates and power generation for batteries. We demonstrate strain engineering in two-dimensional nanomaterials as an approach to manipulate phase transitions and chemical reactivity.

During the intercalation of graphitic materials, lithium atoms are initially distributed randomly throughout the van der Waals (vdW) gaps between carbon sheets (Dilute Stage I), but increasing lithium concentration will trigger a phase transition to ordered structures comprised of intercalated regions vertically separated by unintercalated regions (Fig. 1a). To investigate confinement effects on this process in graphene, we fabricated electrochemical intercalation cells (see Methods and Fig. 1b) using mechanically exfoliated graphene microflakes. The thickness and quality of the exfoliated flakes were assessed with Raman spectroscopy (Extended Data Fig. 1), which confirmed a low defect density[9]. Graphene flakes were assembled into on-chip devices (Extended Data Fig. 2) with metal contacts that function as the cathode of a lithium half-cell. The devices were then placed into an intercalation cell with a lithium/copper anode and immersed in a lithium-containing organic liquid electrolyte for the electrochemical intercalation of Li$^+$ into the vdW gaps of graphene via potentiostatic control (Fig. 1c). Lithium typically initiates staging by forming a Stage IV structure in graphite[10,11] at a concentration around $LiC_{24}$[3,12], where the staging index indicates the number of graphene layers between the intercalant layers. *In situ* Raman spectroscopy of the $E_{2g2}$ mode (G-band) of graphene was used to detect the onset of this staging reaction. As shown in Fig. 1d, the G-band of a several-layer-thick graphene flake with a metal top contact splits into two peaks at 0.7 V vs. Li/Li$^+$ due to the transition from Dilute Stage I to Stage IV. The lower frequency $E_{2g2}(i)$ mode originates from the "interior" unintercalated regions, while the higher frequency $E_{2g2}(b)$ mode originates from the "bounding" layers adjacent to lithium intercalants[11]. *In situ* optical microscopy of the flake during the intercalation shows optical fringes under high applied potential ( < 0.2 V vs. Li/Li$^+$), suggesting the flake began to detach from the substrate surface (Fig. 1e). However, the region adjacent to the metal contact remained anchored to the substrate as the metal contact acted as a mechanical clamp during intercalation.

We systematically investigated lithium staging in clamped microflakes as a function of graphene thickness during intercalation. *In situ* Raman spectroscopy was used to monitor 4-, 10-, 12-, 14-, and 15-layer graphene flakes constrained by two metal electrodes as the potential was dropped from open circuit voltage (OCV) to 0.04 V vs. Li/Li$^+$. The G-band of the 4- and 10-layer flakes split into two peaks at 1.3 V vs. Li/Li$^+$ (Figs. 2a-b), indicating the formation of Stage IV $LiC_{24}$. Additional spectra for the 4-layer flake and a replicate 10-layer device are shown in Extended Data Fig. 3. We observe that the onset of lithium staging is delayed in flakes thicker than 10 layers: the G-peak splits at 0.7 V, 0.14 V, and 0.04 V vs. Li/Li$^+$ for the 12-, 14-, and 15-layer flakes, respectively (Figs. 2c-e). An increase of 1.26 V in the electrochemical potential is required to achieve staging in 15-layer flakes as compared to 4- or 10-layer flakes. Our key finding is that there is a substantial increase in the staging potential for clamped graphene microflakes as a function of thickness, which deviates significantly from the behavior of bulk graphite.

*Post-mortem* analysis of the intercalated flakes with scanning electron microscopy (SEM) shows a thickness-dependent structural change. The optical color change of all flakes post-intercalation can be attributed to an amorphous layer that formed on the surface of the flakes, revealed by SEM (Fig. 3 and Extended Data Fig. 4). X-ray photoelectron spectroscopy shows that the amorphous layer is fluoride and carbonate based (Extended Data Fig. 5), consistent with the formation of a solid-electrolyte interphase (SEI) layer on graphitic electrodes[13]. The 4- and 10-layer graphene flakes show minimal structural change post-intercalation. By stark contrast, the 12-layer flake shows a large structural deformation and interlayer expansion after intercalation. Large cracks radiate out from the electrode (Fig. 3e), while the electrode locally prevents interlayer expansion (Fig. 3f). The 14-layer and 15-layer flakes show increased cracking, SEI growth, and deformation as compared to the 12-layer flake (Extended Data Fig. 4). We note that the thickness-dependent splitting of the G-peak was not caused by the structural damage of the flakes or SEI formation (Extended Data Fig. 6), but was due to the reversible intercalation reaction[11,14]. Thus, concurrent with a delay in lithium staging, thicker graphene experiences greater structural damage.

Based on the observed damage, we conclude that the metal electrodes function as mechanical clamps, causing mechanical stress during intercalation. The electrode locally prevents the intercalation-induced expansion of the vdW gap of graphene flakes, which would generate stress that is heterogeneously distributed across the flake, as shown schematically in Fig. 4a. This would cause the flake to bend and experience in-plane stress ($\sigma$). The bending stiffness (D) of graphene increases as a function of thickness cubed (Fig. 4b)[15,16]. Therefore, thicker flakes would suffer from higher bending-induced mechanical stress during intercalation. Since the bending angle of our graphene flakes is small, we estimate the magnitude of the bending-induced stress by treating them as classical bent plates using continuum mechanics[16]. By modeling multilayer graphene as a clamped rectangular plate (see Supplemental Methods and Extended Data Fig. 7)[17], we calculate that bending-induced uniaxial in-plane stress increases as $t^{2.04}$ and biaxial strain energy (U) increases as $t^{4.08}$, where t denotes the thickness of the graphene flake (Fig. 4c).

The in-plane stress due to bending is estimated to be on the order of $10^4$ Pa based on the modeling (Extended Data Fig. 7). However, the in-plane strain corresponding to this stress is miniscule due to graphene's large Young's modulus of $10^{12}$ Pa [18]. It is also small compared to the expected strain from the 1% expansion of the C-C bonds due to lithium intercalation[19], which causes in-plane stress on the order of $10^9$ Pa in microcrystalline graphite[20,21]. Additionally, while the biaxial strain energy of a 15-layer flake is estimated to be 220 times greater than that of a 4-layer flake due to the increased bending stiffness, the magnitude of the increase is small compared with the observed 1.26 V increase in electrochemical potential required to achieve staging (Extended Data Fig. 8). Therefore, we conclude that the bending-induced strain energy cannot directly explain our *in situ* observations. While device fabrication procedures and the lithium intercalation can lower graphene's mechanical strength, these effects should be similar for all thicknesses of the graphene flakes. In addition, due to graphene's high tolerance of defects[22,23], it would be unlikely that the presence of defects would significantly change the estimated strain. We note that no experimental parameter other than the thickness of the flakes is varied in the intercalation experiments. Thus, the small bending-induced strain energy must be responsible for the observed increase of 1.26 V in electrochemical staging potential.

We propose that the bending-induced strain introduces a nucleation barrier that would modify the kinetics of the phase transformation from Dilute Stage I to Stage IV, affecting the nucleation of Stage IV[24]. The Daumas-Hérold model of lithium staging (Fig. 4d) requires lithium atoms to diffuse throughout the vdW gaps to form ordered lithium domains interspersed laterally throughout the crystal[25]. As the ordered lithium domains move and merge to nucleate Stage IV, there is an elastic energy barrier that contributes to the nucleation barrier ($\Delta G^*$) due to the local increase in mechanical deformation of the graphene layers[26]. We hypothesize that the small bending-induced strain can increase the elastic component of $\Delta G^*$. The rate of nucleation (J) is an exponential function of $\Delta G^*$: to a first approximation $J \propto \exp\left(-\Delta G^*/k_B T\right)$ where $k_B$ and T denote the Boltzmann constant and temperature[27]. Therefore, a small increase in $\Delta G^*$ would lead to an exponentially decreasing nucleation rate, and significantly delay the formation of ordered lithium phases in graphene. Thus, a small bending-induced strain can lead to a large increase in the electrochemical energy required to overcome the nucleation barrier to form Stage IV, which is reflected in the splitting of the G-peak in graphene flakes.

Microscopically, this can also be viewed as a decrease in the diffusion rate of lithium within the vdW gaps of graphene flakes due to the mechanical strain. The bending-induced strain energy would reduce the diffusion rate of lithium, which was theoretically investigated in graphene under in-plane biaxial strain[28]. As the strain energy increases with the thickness of the graphene flakes to the fourth power, the diffusion rate of lithium will decrease more dramatically for thicker flakes. This, again, can be translated into an increase in the $\Delta G^*$ for the nucleation of Stage IV (Fig. 4e). Consequently, the rate of nucleation could be significantly decreased due to mechanical strain, explaining our observed delay in the staging behavior of thicker graphene.

Regardless of the precise mechanism, mechanical strain clearly influences the staging behavior of lithium in graphene. We further confirmed this effect by modulating the strain in 10-layer graphene flakes. In one case, a flake was placed on top of a metal electrode, allowing it to expand freely during intercalation (Extended Data Fig. 9). *In situ* Raman showed the onset of staging at 1.4 V vs. Li/Li$^+$, which is a lower applied potential as compared to the constrained case of the same thickness (1.3 V vs. Li/Li$^+$, Fig. 2b and Extended Data Fig. 3b). In the second case, we observed that a pre-strained 10-layer flake did not initiate staging until 0.12 V vs. Li/Li$^+$ and resulted in increased structural damage compared to other measured 10-layer flakes (Extended Data Fig. 10). Additionally, *in situ* Raman data suggests that some of the pre-loaded strain was relaxed directly preceding the onset of staging at 0.12 V vs. Li/Li$^+$, indicating that it was a key factor in delaying the staging by 1.18 V in this flake. These results indicate that mechanical strain tunes the kinetics of lithium staging within graphene.

Our findings have broad implications for the assembly of nanostructured electrodes for lithium-ion batteries as they suggest even a seemingly insignificant mechanical strain can profoundly affect intercalation kinetics. The effects would be amplified in other layered materials with mechanical strengths lower than graphene, or by increasing intercalant size, which is relevant in sodium and other metal ion-based batteries[29]. The observed effects of strain reach beyond energy storage. Strain engineering has been used to modify the local electronic band structure of two-dimensional (2D) nanomaterials for optoelectronics[30] and catalysis[31], and our results indicate that strain can also be used to modulate the intercalation-induced phase transitions of many 2D materials[32]. Recent interest in the fabrication and intercalation of 2D heterostructures[33–35] has expanded the range of possible 2D nanostructures, and the inherent strain developed due to the lattice mismatch between different materials could potentially be tuned to control intercalation-induced property changes. Tuning the phase transformations of 2D materials with strain has the potential to advance the engineering of devices for applications in logic, optoelectronics, superconductivity, and quantum electronics.

**References:**

1. Armand, M. & Tarascon, J.-M. Building better batteries. *Nature* **451**, 652–657 (2008).
2. Goodenough, J. B. & Park, K. S. The Li-ion rechargeable battery: A perspective. *J. Am. Chem. Soc.* **135**, 1167–1176 (2013).
3. *Graphite Intercalation Compounds I*. (Springer-Verlag, 1990).
4. Safran, S. A. & Hamann, D. R. Electrostatic interactions and staging in graphite intercalation compounds. *Phys. Rev. B* **22**, 606–612 (1980).
5. Kühne, M. *et al.* Reversible superdense ordering of lithium between two graphene sheets. *Nature* **564**, 234–239 (2018).
6. Kühne, M. *et al.* Ultrafast lithium diffusion in bilayer graphene. *Nat. Nanotechnol.* **12**, 895–900 (2017).
7. Bao, W. *et al.* Approaching the limits of transparency and conductivity in graphitic materials through lithium intercalation. *Nat. Commun.* **5**, 4224 (2014).
8. Xiong, F. *et al.* Li intercalation in MoS2: in situ observation of its dynamics and tuning optical and electrical properties. *Nano Lett.* **15**, 6777–6784 (2015).
9. Ferrari, A. C. & Basko, D. M. Raman spectroscopy as a versatile tool for studying the properties of graphene. *Nat. Nanotechnol.* **8**, 235–246 (2013).
10. Dahn, J. R., Fong, R. & Spoon, M. J. Suppression of staging in lithium-intercalated carbon by disorder in the host. *Phys. Rev. B* **42**, 6424–6432 (1990).
11. Inaba, M. *et al.* In situ Raman study on electrochemical Li intercalation into graphite. *J. Electrochem. Soc.* **142**, 20–26 (1995).
12. Fischer, J. E., Fuerst, C. D. & Woo, K. C. Staging transitions in intercalated graphite. *Synth. Met.* **7**, 1–12 (1983).
13. Heiskanen, S. K., Kim, J. & Lucht, B. L. Generation and evolution of the solid electrolyte interphase of Lithium-Ion batteries. *Joule* **3**, 2322–2333 (2019).
14. Sole, C., Drewett, N. E. & Hardwick, L. J. In situ Raman study of lithium-ion intercalation into microcrystalline graphite. *Faraday Discuss.* **172**, 223–237 (2014).
15. Zhang, D. B., Akatyeva, E. & Dumitric, T. Bending ultrathin graphene at the margins of continuum mechanics. *Phys. Rev. Lett.* **106**, 3–6 (2011).
16. Han, E. *et al.* Ultrasoft slip-mediated bending in few-layer graphene. *Nat. Mater.* **19**, 305–209 (2019).
17. Timoshenko, S. & Woinowsky-Krieger, S. *Theory of Plates and Shells*. (McGraw-Hill, 1959).
18. Lee, C., Wei, X., Kysar, J. W. & Hone, J. Measurement of the elastic properties and intrinsic strength of monolayer graphene. *Science* **321**, 385–388 (2008).
19. Qi, Y., Guo, H., Hector, L. G. & Timmons, A. Threefold increase in the Young's modulus of graphite negative electrode during lithium intercalation. *J. Electrochem. Soc.* **157**, A558 (2010).

20. Zou, J., Sole, C., Drewett, N. E., Velický, M. & Hardwick, L. J. In situ study of Li intercalation into highly crystalline graphitic flakes of varying thicknesses. *J. Phys. Chem. Lett.* **7**, 4291–4296 (2016).

21. Xie, H. *et al.* In situ measurement of rate-dependent strain/stress evolution and mechanism exploration in graphene electrodes during electrochemical process. *Carbon* **144**, 342–350 (2019).

22. Lee, G. H. *et al.* High-strength chemical-vapor-deposited graphene and grain boundaries. *Science* **340**, 1074–1076 (2013).

23. Zandiatashbar, A. *et al.* Effect of defects on the intrinsic strength and stiffness of graphene. *Nat. Commun.* **5**, 3186 (2014).

24. Funabiki, A., Inaba, M., Abe, T. & Ogumi, Z. Nucleation and phase-boundary movement upon stage transformation in lithium-graphite intercalation compounds. *Electrochim. Acta* **45**, 865–871 (1999).

25. Daumas, N. & Herold, A. Relations between phase concept and reaction mechanics in graphite insertion compounds. *C. R. Acad. Sci* **268**, 373 (1969).

26. Ulloa, S. E. & Kirczenow, G. Nonlinear theory of domain walls and domain effective interactions in intercalation compounds. *Phys. Rev. B* **33**, 1360–1371 (1986).

27. Karthika, S., Radhakrishnan, T. K. & Kalaichelvi, P. A review of classical and nonclassical nucleation theories. *Cryst. Growth Des.* **16**, 6663–6681 (2016).

28. Hao, F. & Chen, X. First-principles study of lithium adsorption and diffusion on graphene: The effects of strain. *Mater. Res. Express* **2**, 105016 (2015).

29. Choi, J. W. & Aurbach, D. Promise and reality of post-lithium-ion batteries with high energy densities. *Nat. Rev. Mater.* **1**, 16013 (2016).

30. Iff, O. *et al.* Strain-tunable single photon sources in WSe2 monolayers. *Nano Lett.* **19**, 6931–6936 (2019).

31. Li, H. *et al.* Activating and optimizing MoS2 basal planes for hydrogen evolution through the formation of strained sulphur vacancies. *Nat. Mater.* **15**, 48–53 (2016).

32. Yazdani, S., Yarali, M. & Cha, J. J. Recent progress on in situ characterizations of electrochemically intercalated transition metal dichalcogenides. *Nano Res.* **12**, 2126–2139 (2019).

33. Geim, A. K. & Grigorieva, I. V. Van der Waals heterostructures. *Nature* **499**, 419–425 (2013).

34. Deng, D. *et al.* Catalysis with two-dimensional materials and their heterostructures. *Nat. Nanotechnol.* **11**, 218–230 (2016).

35. Bediako, D. K. *et al.* Heterointerface effects in the electrointercalation of van der Waals heterostructures. *Nature* **558**, 425–429 (2018).

**Figures:**

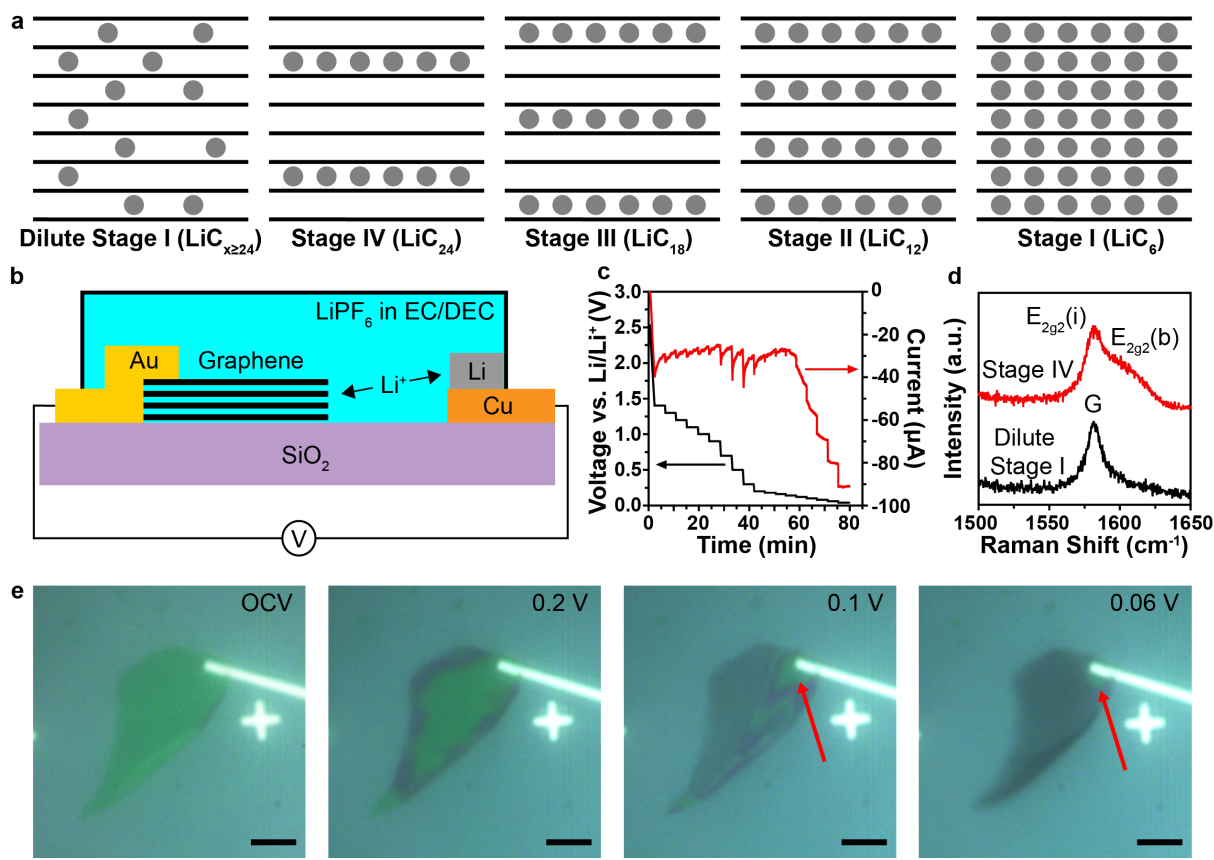

**Fig. 1 | Electrochemical intercalation of Li$^+$ into graphene. a,** Cross-section depiction of the intercalation of lithium (gray circles) between graphene sheets progressing from Dilute Stage I to fully-lithiated Stage I. **b,** Schematic of an *in situ* electrochemical intercalation cell. **c,** Representative voltage and current profiles during the potentiostatic intercalation of graphene flakes. **d,** *In situ* Raman spectra of several-layer-thick graphene during lithiation. The G-peak of Dilute Stage I at 0.9 V vs. Li/Li$^+$ splits into the E$_{2g2}$(i) and E$_{2g2}$(b) modes of Stage IV at 0.7 V vs. Li/Li$^+$. **e,** Optical images of the flake from (d) taken at open circuit voltage (OCV) and under decreasing potential vs. Li/Li$^+$, scale bars 10 μm. The series shows the gradual detachment of the flake from the substrate, while the red arrows indicate that the regions adjacent to the electrode remain attached.

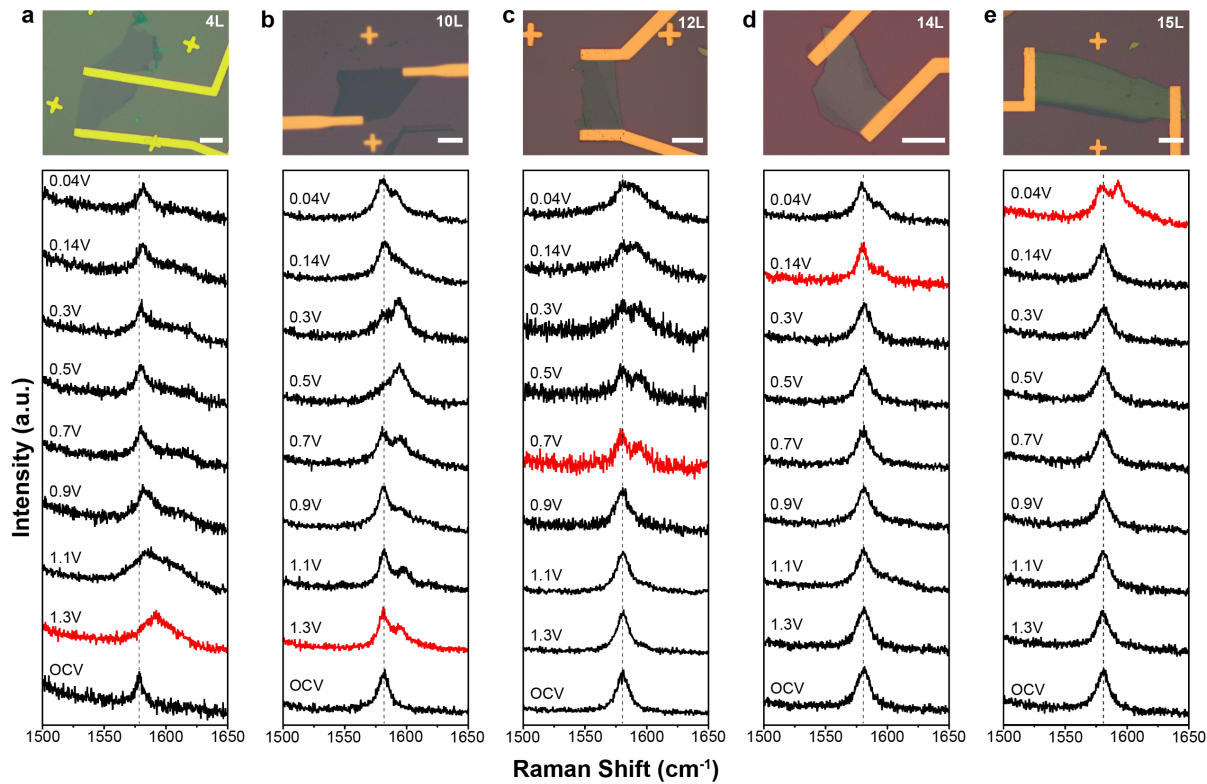

**Fig. 2 | Thickness-dependent delay in lithium staging of graphene clamped by metal contacts.** Optical images (top) and *in situ* Raman spectra (bottom) taken during the intercalation of 4-, 10-, 12-, 14-, and 15-layer graphene flakes in panels **a**, **b**, **c**, **d**, and **e**, respectively. The graphene flakes are clamped by two metal top-contacts, which are used for Li intercalation (scale bars 10 μm). *In situ* Raman spectra of the graphene G-band were taken at various potentials versus Li/Li$^+$. The red spectra indicate the onset of lithium staging within the flakes. The dashed line indicates the G-band position at OCV.

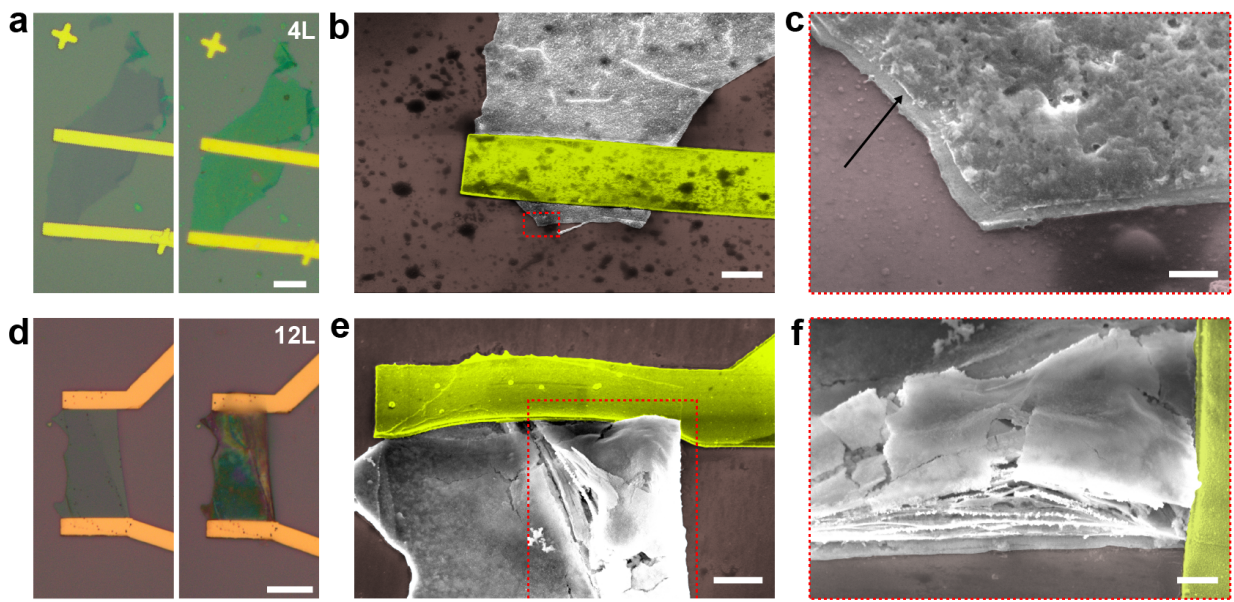

**Fig. 3 | Structural changes to graphene clamped by metal contacts after intercalation.**
**a,** Optical images of 4-layer graphene before (left) and after (right) intercalation, scale bar 10 μm. **b,** SEM micrograph of the flake in (a), scale bar 2 μm. **c,** SEM micrograph of the red-dashed area in (b), scale bar 200 nm. The arrow indicates the interface between the flake and the SEI. **d,** Optical images of 12-layer graphene before (left) and after (right) intercalation, scale bar 10 μm. **e,** SEM micrograph of the flake in (d), scale bar 2 μm. **f,** SEM micrograph of the red-dashed area in (e), scale bar 1 μm. The SEM micrographs were taken at a 40º tilt-angle, with the gold contact and substrate false-colored yellow and purple, respectively.

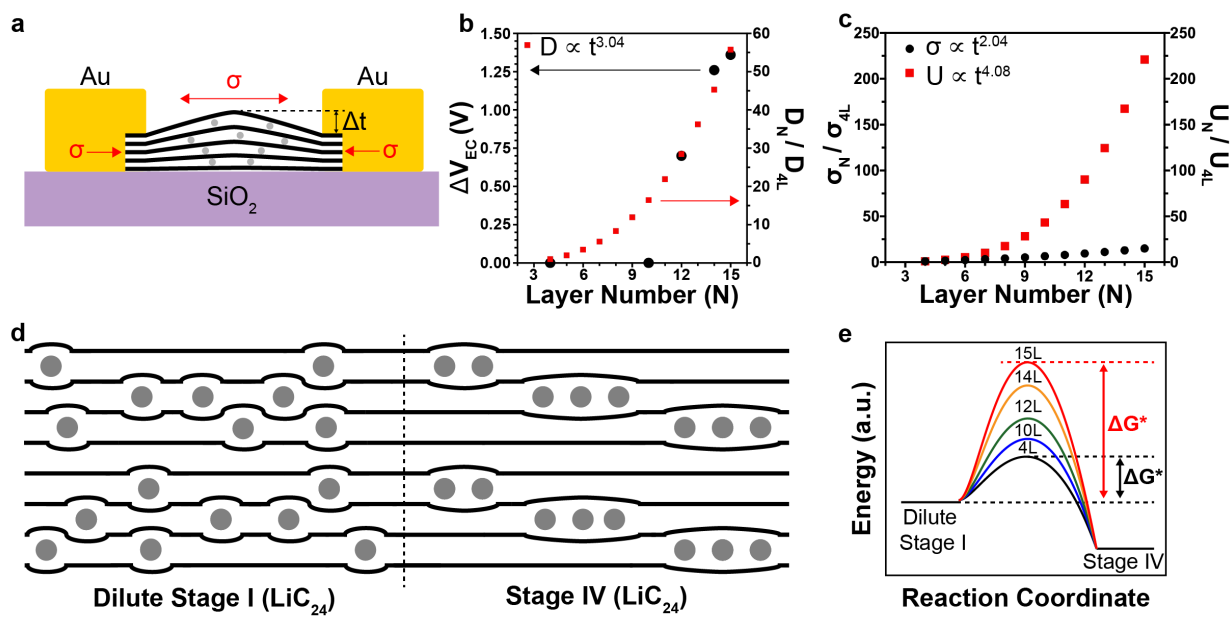

**Fig. 4 | Effects of mechanical stress induced by metal contacts on the kinetics of lithium staging. a,** Cross-sectional schematic of lithium-intercalated graphene with two top contacts (axes not to scale). Li$^+$ (gray circles) induces an increase in thickness ($\Delta t$), causing in-plane stress ($\sigma$). **b,** The change in electrochemical voltage ($\Delta V_{EC}$) required for staging as function of layer number (circles) and the relative bending stiffness (D) of N-layer graphene (squares), normalized to 4-layer graphene. **c,** Relative increase in uniaxial stress ($\sigma$) and biaxial strain energy (U) predicted by continuum mechanics for N-layer graphene normalized to 4-layer graphene. **d,** Schematic of the phase transformation from Dilute Stage I to Stage IV at the same lithium concentration of LiC$_{24}$ according to the Daumas-Hérold model of lithium staging. **e,** Schematic reaction coordinate of lithium staging (energy axis not to scale). The nucleation barrier ($\Delta G^*$) is shown for 4- (black) and 15-layer (red) graphene.

**Materials and Methods**

Device Fabrication and Assembly of Intercalation Cell

Multilayer graphene flakes were exfoliated from bulk graphite (NGS Naturgraphit GmbH) using the scotch-tape method onto $SiO_2$/Si substrates. The substrates were cleaned via sonication in acetone and isopropyl alcohol, and treated with $O_2$ plasma prior to exfoliation. Suitable flakes were identified via optical microscopy and their thickness was determined via Raman spectroscopy using the shape of the 2D peak and the intensity of the Si substrate through the flake as shown in Extended Data Fig. 1 and described previously[36,37].

Selected flakes were transferred to a new substrate using the KOH-assisted transfer technique outlined in Extended Data Fig. 2. Briefly, a droplet of epoxy (Scotch-Weld) was allowed to cure on a glass slide, forming a hemispherical drop approximately 0.5 mm in diameter. A 13% by weight solution of polypropylene carbonate (PPC, Sigma Aldrich) dissolved in anisole (Sigma Aldrich) was spun onto the epoxy drop at 3000 RPM for 2 minutes. The PPC was cured for 2 minutes at 90ºC and then allowed to cool. A thin layer of epoxy was then spread around the edge of the PPC, sealing it onto the glass slide. Using a transfer stage, the droplet was aligned above the graphene flake of interest. The glass slide was lowered until the PPC contacted the graphene flake. About 20 μL of a 2M aqueous solution of potassium hydroxide (KOH, Sigma Aldrich) was added to the substrate, instantaneously etching the top few Å of $SiO_2$, and releasing the graphene flake. The glass slide was then lifted up with the graphene flake attached to the PPC. The slide was washed in deionized water and allowed to dry. The flake was then transferred to a target $SiO_2$/Si substrate with alignment marks for lithography by lowering the PPC/graphene down to make contact on the desired area of the target substrate. The PPC was melted by heating the substrate to 110ºC for 5 minutes. The glass slide was then lifted, leaving the flake attached to the new substrate, and covered with melted PPC. The PPC was subsequently dissolved overnight in chloroform, leaving the transferred flake on the target substrate.

Electrodes were then patterned onto the flakes using SEM-based electron beam lithography (Nabity NPGS, Helios G4 FIB-SEM) and 100 nm gold contacts were deposited with a 10 nm Cr wetting layer using thermal evaporation (Mbraun EcoVap). Some devices were fabricated with bottom contacts by first depositing 10 nm Cr / 100 nm Au onto a $SiO_2$/Si substrate, and then transferring flakes directly onto the contact using the transfer technique described above.

Fabricated devices were then placed in a custom liquid intercalation cell as shown in Extended Data Fig. 2g. The cell consists of a glass dish set into a polytetrafluoroethylene (PTFE) base, with copper and aluminum contact pads attached to the glass insert with an epoxy adhesive (Scotch-Weld). These leads run to the exterior of the cell and are sealed off with epoxy. The substrate containing the graphene devices was attached to the glass insert with double-sided copper tape, and then wire-bonded to the aluminum contact pads with aluminum wire. Before adding the electrolyte, a Raman spectrum of the pristine flake was taken.

The intercalation cell was then placed inside an argon glovebox, where a small piece (~ 1 $mm^2$) of lithium metal (Sigma Aldrich) was attached to the copper contact pads using copper tape. The electrolyte, a battery-grade solution of 1 M lithium hexafluorophosphate ($LiPF_6$) in 50/50 v/v ethylene carbonate / diethyl carbonate (Sigma Aldrich), was added to the cell so that both the lithium and graphene flake were submerged. The PTFE base was then bolted in between two stainless steel plates, forming an airtight seal against a fluorinated ethylene propylene (FEP)-encapsulated viton o-ring. The top plate was fitted with a window made of optical glass, allowing for the graphene flake to be viewed with an optical microscope. The cell was then brought out of the glovebox for immediate measurement.

Electrochemical Intercalation

The intercalation cell was connected to a Biological SP300 workstation for the electrochemical intercalation of $Li^+$. Typical open circuit voltages (OCV) were about 2.5-2.7 V vs. $Li/Li^+$. Before intercalation, a Raman spectrum was taken at OCV. No change in peak shape was observed after immersion in electrolyte, and no contaminating peaks were observed in the region of the G-peak (Extended Data Fig. 2h). Lithium was intercalated into graphene potentiostatically using successive voltage steps from OCV down to 0.04 V. The cell was held at OCV for 30 seconds before dropping the potential to 1.4 V vs. $Li/Li^+$ at a scan rate of 10 $mVs^{-1}$. The cell was held at 1.4 V vs. $Li/Li^+$ while a Raman spectrum of the flake was taken. The cell potential was subsequently dropped to 1.3 V, 1.2 V, 1.1 V, 1.0 V, 0.9 V, 0.7 V, 0.5 V, 0.3 V, 0.2 V, 0.18 V, 0.16 V, 0.14 V, 0.12 V, 0.10 V, 0.08 V, 0.06 V, and 0.04 V vs. $Li/Li^+$ at a rate of 10 $mVs^{-1}$. The cell was held at each potential for 4 minutes and a Raman spectrum was taken at each step after fixing the potential for 1 minute. After removing the intercalation potential, the flakes were allowed to rest for approximately 10 minutes.

Raman Characterization

All Raman spectra were taken with a Horiba LabRAM HR Evolution Spectrometer using a 633 nm HeNe laser with an 1800 lines/mm diffraction grating. Before intercalation, all samples were characterized at a laser power of ~3 mW to avoid damage to the flakes. Spectra of the G and 2D peaks were collected with a 100x optical lens; however, quantitative analysis of flake thickness via the Si peak was conducted with a 50x long-working distance lens with a numerical aperture of 0.5. Once devices were immersed in the liquid electrolyte inside the intercalation cell, spectra were taken using the 50x long-working distance lens. Due to scattering by the liquid electrolyte, a laser power of ~7.5 mW was used to increase the signal-to-noise ratio. *In situ* spectra at each potential were collected with twenty 5-second exposures (Fig. 2 and Extended Data Fig. 3). Post-intercalation, the devices were taken out of the intercalation cell and Raman spectra were collected at a laser power of ~3mW with a 100x lens (Extended Data Fig. 6).

*Post-Mortem* Characterizations

Post-intercalation, the cell was disassembled and the substrate with the graphene device was removed from the electrolyte and placed into an isopropyl alcohol wash. The device was then immediately characterized with optical microscopy and Raman spectroscopy. Devices were examined with SEM (Helios G4 FIB-SEM) at a tilt angle of both 0 and 40 degrees (Fig. 3 and Extended Data Fig. 4). XPS analysis (Extended Data Fig. 5) was conducted with a monochromatic 1486.7 eV Al Kα x-ray source on a PHI VersaProbe II x-ray photoelectron spectrometer with a 0.47 eV system resolution. The energy scale was calibrated using Cu $2p_{3/2}$ (932.67 eV) and Au $4f_{7/2}$ (84.00 eV) peaks on a clean copper plate and clean gold foil. An x-ray beam with a diameter of 20 μm was directed onto the areas of interest using x-ray induced secondary electron imaging (SXI). XPS spectra were normalized using the Si 2p peak from the $SiO_2$ substrate at 103.3 eV as an internal standard.

**Methods and Extended Data References:**

36. Ferrari, A. C. *et al.* Raman spectrum of graphene and graphene layers. *Phys. Rev. Lett.* **97**, 187401 (2006).

37. Li, X. L. *et al.* Layer number identification of intrinsic and defective multilayered graphenes up to 100 layers by the Raman mode intensity from substrates. *Nanoscale* **7**, 8135–8141 (2015).

38. B. Zhang *et al.*, Role of 1,3-propane sultone and vinylene carbonate in solid electrolyte interface formation and gas generation. *J. Phys. Chem. C*. **119**, 11337–11348 (2015).

39. B. Ziv *et al.*, Investigation of the reasons for capacity fading in Li-ion battery cells. *J. Electrochem. Soc.* **161**, A1672–A1680 (2014).

40. A. M. Andersson *et al.*, Surface characterization of electrodes from high power lithium-ion batteries. *J. Electrochem. Soc.* **149**, A1358 (2002).

41. Y. Taki, O. Takai, XPS structural characterization of hydrogenated amorphous carbon thin films prepared by shielded arc ion plating. *Thin Solid Films*. **316**, 45–50 (1998).

42. M. Huang *et al.*, Phonon softening and crystallographic orientation of strained graphene studied by Raman spectroscopy. *Proc. Natl. Acad. Sci.* **106**, 7304–7308 (2009).

43. D. Yoon, Y. W. Son, H. Cheong, Strain-dependent splitting of the double-resonance Raman scattering band in graphene. *Phys. Rev. Lett.* **106**, 155502 (2011).

**Acknowledgments:** The authors thank V. Ozolins, H. Wang, and E. Altman for their discussion of this work. Device fabrication and characterization was carried out at the Yale West Campus Materials Characterization Core and the Yale West Campus Clean Room with support from M. Li and L. Wang. *In situ* cell fabrication was assisted by V. Bernardo at the Yale J.W. Gibbs Professional Shop and by D. Smith at the Yale Scientific Glassblowing Laboratory. This work was supported by the Army Research Office (W911NF-18-1-0367). J.V.P. was supported by the National Defense Science and Engineering Graduate (NDSEG) Fellowship Program, sponsored by the Air Force Research Laboratory (AFRL), the Office of Naval Research (ONR), and the Army Research Office (ARO). The intercalation cell was developed with support from the National Science Foundation (CAREER CBET No. 1749742).

**Author contributions:** J.V.P. and J.J.C. conceived the experiments. J.V.P. fabricated and characterized the devices, and analyzed the data. S.Y., D.J.H., and J.V.P. designed the custom intercalation cell. S.Y. and M.Y. contributed to the development of electrochemical and fabrication methods. S.N.R. assisted with graphene characterization. J.V.P. and J.J.C. wrote the manuscript with input from all authors.

**Competing interests:** The authors declare no competing financial interest.

**Additional Information:**

**Extended Data** is available for this paper.

**Supplemental Methods** are available for this paper.

**Correspondence and requests for materials** should be addressed to J.J.C.

**Reprints and permissions information** is available at www.nature.com/reprints